\documentclass[10pt,twocolumn,twoside,journal]{IEEEtran}
\IEEEoverridecommandlockouts
\usepackage{amsmath,graphicx,amssymb,mathtools,bm}
\usepackage{subfigure}
\usepackage{hyperref}
\usepackage{cite}
\usepackage{amsfonts}
\usepackage{xcolor,color}
\usepackage{float}
\usepackage{verbatim}   
\usepackage{mathrsfs} 
\usepackage{algorithm} 
\usepackage{algorithmic} 
\usepackage{booktabs}
\usepackage{textcomp}  
\usepackage{multirow}  
\usepackage{lettrine}    
\usepackage{stfloats}
\usepackage{caption,subcaption}
\UseRawInputEncoding
\usepackage[linesnumbered,ruled,vlined,algo2e]{algorithm2e}

\begin{document}
\newcommand{\tabincell}[2]{\begin{tabular}{@{}#1@{}}#2\end{tabular}}

\title{Movable Antenna-Enhanced Wireless Communications: General Architectures and Implementation Methods}
\author{Boyu Ning, Songjie Yang, Yafei Wu, Peilan Wang, Weidong Mei, Chau Yuen, and Emil Bj{\"o}rnson
\thanks{B. Ning, S. Yang, P. Wang, and W. Mei {\it (corresponding author)} are with the National Key Laboratory of Wireless Communications, University of Electronic Science and Technology of China, Chengdu 611731, China (e-mail: boydning@outlook.com, yangsongjie@std.uestc.edu.cn, peilan\_wangle@uestc.edu.cn, wmei@uestc.edu.cn).}
\thanks{Y. Wu {\it (corresponding author)} is with the EHF Key Laboratory of Fundamental Science, School of Electronic Science and Engineering, University of Electronic Science and Technology of China, Chengdu 611731, China (e-mail: wuyafei@uestc.edu.cn).}
\thanks{C. Yuen is with the School of Electrical and Electronics Engineering, Nanyang Technological University, Singapore 639798 (email: chau.yuen@ntu.edu.sg).}
\thanks{Emil Bj{\"o}rnson is with Department of Computer Science, KTH Royal Institute of Technology, Stockholm, Sweden (email: emilbjo@kth.se).}
}
\maketitle
\begin{abstract}
Movable antennas (MAs), traditionally explored in antenna design, have recently garnered significant attention in wireless communications due to their ability to dynamically adjust the antenna positions to changes in the propagation environment. However, previous research has primarily focused on characterizing the performance limits of various MA-assisted wireless communication systems, with less emphasis on their practical implementation. To address this gap, in this article, we propose several general MA architectures that extend existing designs by varying several key aspects to cater to different application scenarios and tradeoffs between cost and performance. Additionally, we draw from fields such as antenna design and mechanical control to provide an overview of candidate implementation methods for the proposed MA architectures, utilizing either direct mechanical or equivalent electronic control. Simulation results are finally presented to support our discussion.
\end{abstract}

\section{Introduction} 
The past few decades have witnessed the wide adoption of multi-antenna technologies in wireless communication systems. Thanks to their high beamforming, spatial multiplexing, and diversity gains, multi-antenna technologies can significantly boost the capacity and reliability of wireless communications. With the explosive growth in demand for wireless connectivity and data rates, massive and even extremely large-scale antenna arrays are anticipated to gain wider adoption in the deployment of next-generation wireless communication systems. However, this also results in a dramatic increase in the antenna cost and radio-frequency (RF) energy consumption, which pose significant challenges for achieving green and sustainable wireless communications. Furthermore, despite the large antenna array, the current multi-antenna technologies are limited to adapting to wireless channels without the capability to actively reshape them, which thus may result in suboptimal performance.

\begin{figure*}
\centering
\includegraphics[width = 1.01\textwidth]{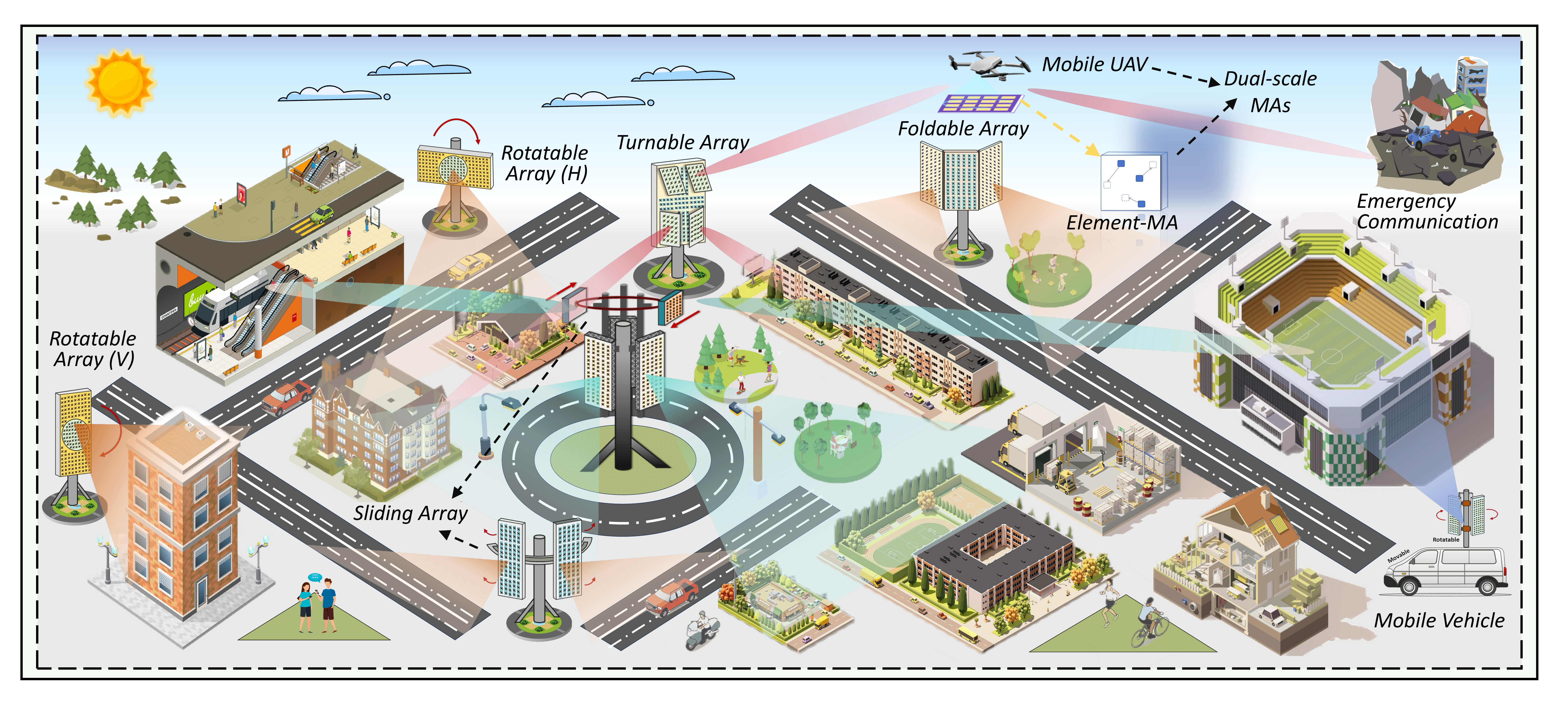} 
\caption{Promising applications of MAs, where the movement can be based on rotating, sliding, folding, or a mix of these things.} \label{application}
\label{MA_FIG}
\end{figure*}

To overcome this limitation, movable antenna (MA) technologies have obtained increasing attention in wireless communications \cite{SA2}. Compared to the conventional fixed-position antennas (FPAs), MAs can proactively reshape the wireless channels into a more favorable condition for data transmission by circumventing the positions/angles that may experience deep fading for desired users and/or strong interference with undesired users, even without the need for multiple antennas \cite{SA2,FAS1,XT1,XT2}. Moreover, in multiple-input multiple-output (MIMO) systems, multiple MAs can achieve higher array and spatial multiplexing gains by optimizing their positions/rotations to reshape MIMO channels. Furthermore, the geometry of an MA array can be dynamically reconfigured based on the user distribution for various purposes, e.g., interference nulling and multi-beam coverage, by jointly optimizing multiple MAs. Last but not least, the time scale of antenna movement can be adapted to various scenarios, including short-term small-scale fading, average performance under fading for currently scheduled users, and even longer-term adaptation to changes in the geographic user distribution. As such, it is anticipated that MAs can find a broad range of applications in the current wireless communication systems to enhance the coverage and communication performance, as shown in Fig.\,\ref{application}.

However, despite the recent theoretical advancements in MAs, the research is still in its infancy. In particular, current studies primarily focus on algorithmic design and characterizing the performance limit of MA-assisted wireless communication systems under various setups by optimizing the positions of antenna elements, assuming that each element's position can be individually and globally adjusted within the transmit/receive region \cite{EMA1,EMA2,EMA3,XT3,XT4,XT5}. However, this architecture may lead to prohibitively high energy consumption and position-tuning delays, especially for massive antenna arrays, thus challenging the practical implementation of MAs. Furthermore, the current MA architecture requires a fundamental paradigm shift from existing FPA arrays, potentially resulting in high network infrastructure costs and incompatibility with existing communication protocols. Therefore, an urgent need arises to explore new and cost-effective MA architectures to better balance communication performance and implementation complexity. On the other hand, the practical applications of any MA architecture are fundamentally limited by their implementation methods. Excessively high implementation difficulty can negate the potential performance gains from antenna position optimization. However, there are still significant knowledge gaps in wireless communications regarding the development of low-cost and low-latency implementation methods that meet the requirements of practical communication systems.

To address these key challenges for MAs, this article proposes several innovative architectures and practical implementation methods to pave the way for their efficient integration into current/future wireless communication systems. In the rest of this article, we first introduce the proposed MA architectures, which extend existing designs by varying key aspects such as the moving unit (element versus array), flexibility (position-tuning versus joint position and rotation tuning), scale (small- versus large-scale movement), range (local versus global movement), and functionality (fully versus partially MAs). These variations are tailored to specific and typical communication scenarios through the sliding, turning, and/or folding of the antenna elements/arrays, as shown in Fig.\,\ref{application}. Next, we draw from fields such as antenna design and mechanical control to provide an overview of candidate implementation methods for MAs, which utilize either direct mechanical control or equivalent electronic control. We compare their respective advantages and disadvantages and discuss their preferred applications in wireless communications. Finally, numerical examples are provided to demonstrate the effectiveness of the proposed MA architectures and implementation methods. It is worth noting that although MAs and similar technologies (e.g., fluid antennas and flexible-position MIMO) have been reviewed in prior articles \cite{SA2,FAS2,FPA}, a dedicated discussion on their general architectures and practical implementation methods is still lacking in the literature to the best of our knowledge, which motivates this work.

\section{General Architectures with MAs}
In this section, we first present the common architectures with antenna {\it element} movement and then propose more efficient designs for antenna {\it array} movement tailored to various scenarios. Additionally, we introduce a general model for MAs incorporating both small- and large-scale movements to enhance channel reconfiguration capabilities.

\subsection{Element-Level MAs}
\begin{figure}
\centering
\includegraphics[width = 0.48\textwidth]{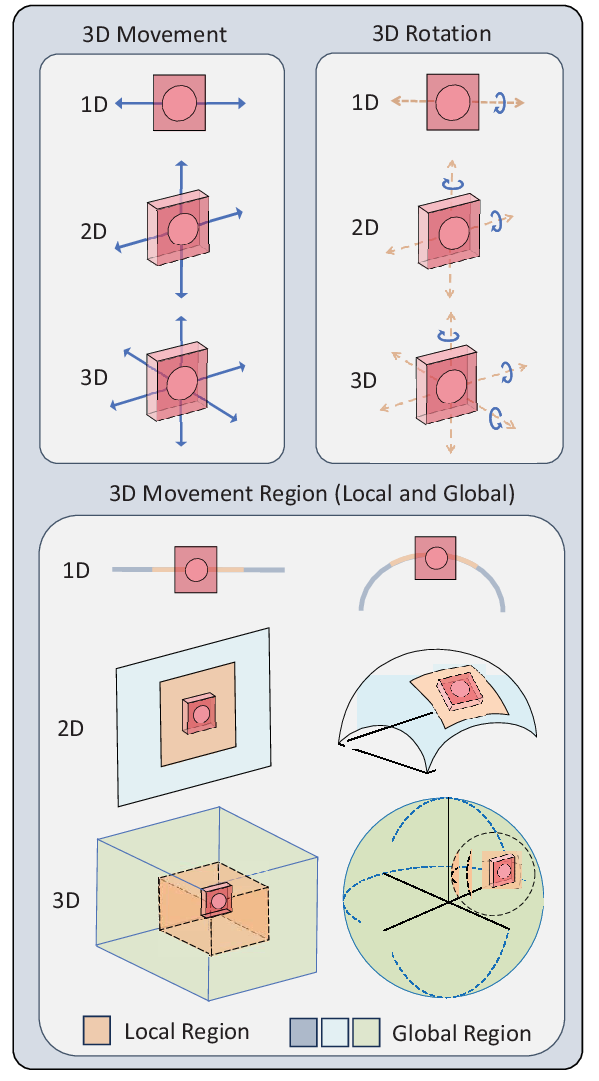} 
\caption{Illustration of element-level MAs} 
\label{EM}
\end{figure}
One straightforward MA architecture involves moving each individual antenna {\it element} within a given region at the transmitter or receiver. For example, as shown in Fig. \ref{EM}, an MA element can be moved within a one-dimensional (1D) array, a two-dimensional (2D) region, or even a three-dimensional (3D) space. Furthermore, existing studies have mostly focused on optimizing performance for MAs within regular regions, such as linear arrays or planar areas. However, in certain scenarios, regions of special shape may also be utilized for antenna movement, e.g., the curved lines or surfaces shown in Fig. \ref{EM}. Compared to regular-shaped regions, special-shaped regions allow MAs to exploit channel variations over a larger space. For example, in 1D movement, a linear array only changes the antenna's location in one dimension, while a curved array can alter its location in more dimensions, which is particularly useful for serving users with a scattered distribution.

In addition to adjusting the location of an MA element, its orientation can also be modified in a 3D space to achieve different yaw, roll, and pitch angles, as shown in Fig. \ref{EM}, which can affect the antennas’ radiation pattern and polarization. For example, with a directional MA element, altering its orientation can change the direction of its main lobe, enabling efficient beam tracking to serve mobile users. This capability also allows for more flexible beam coverage over time if user distribution varies within a given area. Furthermore, with polarized antennas at both the transmitter and receiver, antenna rotation can adjust the polarization direction of the transmitted and received signals, achieving more efficient polarization matching without the need for decomposition and reconstruction operations. It is worth noting that the orientation of the MA elements extends the concept of remote electrical tilt (RET) antennas in 3GPP, which allows the vertical tilt angle of base station (BS) array to be controlled remotely, to a broader range of applications in wireless communications. By combining the capabilities of MAs for both 3D orientation and 3D movement, each MA has six degrees-of-freedom (DoFs) in total.


Nonetheless, adjusting the positions and rotational angles of multiple MA elements simultaneously at a rapid time scale (e.g., for adapting to small-scale fading) can entail high implementation complexity and power consumption. Additionally, this may result in considerably high overhead in channel estimation and position/orientation optimization due to the large feasible region for 3D movement. Therefore, instead of altering the location of each MA within the entire 3D space, a more viable strategy is to restrict its movement within a local region. For example, in a 1D linear array, the array can be uniformly divided into multiple subregions, and the location of each MA element can only be altered within its designated subregion. Moreover, in certain scenarios, such as terrestrial communications, the feasible space for the orientation angle of each MA can be restricted as well due to the generally much lower altitudes of terrestrial users compared to BSs. As such, the main lobe of the BS antennas should be pointed downwards to serve these users effectively. In general, there is a fundamental trade-off between enhancing the performance of the MA-assisted system and reducing the complexity of antenna movement. Properly resolving this trade-off based on specific system setups is a crucial problem worthy of in-depth investigation, as will be pursed next by designing new array-level MA architectures.

\begin{figure*}
\centering
\includegraphics[width = 0.98\textwidth]{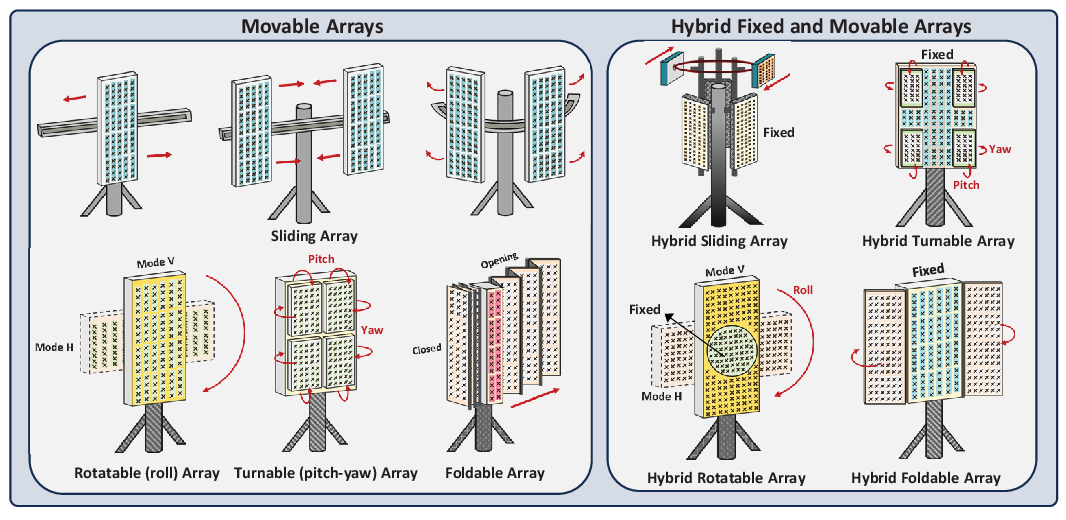} 
\caption{Illustration of array-level MAs} 
\label{MA_FIG}
\end{figure*}
\subsection{Array-Level MAs}
Unlike element-level MAs, where each antenna element can move or rotate independently, we introduce an alternative architecture in this subsection, namely, array-level MAs. This approach divides the entire antenna array into multiple sub-arrays with all antenna elements within each sub-array moving or rotating in unison, while the movement or rotation of antenna elements can vary between different sub-arrays. Notably, the array-level MA architecture is more amenable to practical hardware implementation than its element-level counterpart, as it reduces the complexity and power consumption associated with independently controlling each MA element. In addition, it allows for the use of existing compact antenna arrays unlike element-level MAs that require unusually large spacing between the elements to have space for movements. In the following, we propose various forms of array-level MAs and elucidate their specific advantages, including sliding array, rotatable array, turnable array, and foldable array, as illustrated in Fig. \ref{MA_FIG}.

\subsubsection{Sliding array} First, a sliding array is an array of movable sub-arrays each able to flexibly change its location within a given region. In the example depicted in Fig. \ref{MA_FIG}, the sliding array consists of two movable sub-arrays, each capable of sliding along a 1D straight or curved line. The positions of these sub-arrays can be flexibly adjusted in favor of wireless transmission/reception. Compared to element-level MAs, the sliding array offers less flexibility for channel reconfiguration but more efficient implementation by eliminating the need for individual location control of each antenna element. Moreover, by properly adjusting the position of each sub-array, the deep-fading and/or high-interference positions can still be avoided, thereby ensuring the communication performance for the entire array. In addition, by reducing/increasing the distances between any two adjacent sub-arrays, more flexible beam coverage can be achieved to cover centralized/separated regions, each associated with and covered by a sub-array. Last but not least, increasing the distance among the sub-arrays can more efficiently enhance the aperture of the entire antenna array, thereby proactively creating a near-field propagation environment. This enables beamfocusing where signals are steered both in angle and distance and also provides the ability to send multiple streams to a single user even for line-of-sight (LoS) propagation\cite{Emil}.

\subsubsection{Rotatable array} Second, a rotatable array refers to an antenna array that can rotate along its roll angle (not necessarily the entire 3D space) to achieve adaptive wireless coverage. For a given number of antennas, the aspect ratio of a rotatable array determines its coverage capabilities in the horizontal and vertical directions. A larger number of antennas in the horizontal/vertical dimension enhances the spatial resolution in its respective direction. As illustrated in Fig. \ref{MA_FIG}, the rotatable array can operate in two modes, i.e., Mode H (horizontal) and Mode V (vertical). If the users are evenly distributed along the horizontal (or vertical) dimension, Mode H (or Mode V) should be utilized to ensure adequate resolution for their separation. In particular, in urban scenarios with high-rise buildings, the user distribution may vary throughout the day. For example, during peak commuting hours, users are predominantly at ground level on the streets, necessitating broader horizontal coverage. Conversely, during working hours, users typically reside in high-rise buildings, requiring more substantial vertical coverage. In this context, the rotatable array can dynamically adjust its angle over a large time scale to ensure the maximum coverage.

\subsubsection{Turnable array} Third, a turnable array consists of multiple sub-arrays, each capable of rotating along its yaw and pitch angles. Unlike a rotatable array, these sub-arrays cannot rotate around their roll angles due to their close proximity to adjacent sub-arrays. In the example shown in Fig. \ref{MA_FIG}, the turnable array comprises four sub-arrays. The turnable array is particularly advantageous in scenarios with distributed user clusters, as each sub-array can be rotated to serve a specific user cluster. Compared to traditional FPAs, a turnable array can better balance the beamforming gain among user clusters in different directions. Furthermore, different numbers of sub-arrays can be allocated to each user cluster based on their rate requirements. As such, it is crucial to optimize the associations between sub-arrays and user clusters to ensure the overall communication performance, given the finite number of sub-arrays.

\subsubsection{Foldable array} Finally, a foldable array comprises multiple sub-arrays capable of adaptive folding or unfolding into specific geometric shapes to achieve desired antenna properties, akin to origami. Notably, the size of antenna arrays in practical BSs is often limited by wind resistance. However, foldable arrays offer a flexible solution to this issue: the sub-arrays can be folded when it is windy and unfolded when the wind subsides. Therefore, the foldable array should be equipped with anemometers, and its use should take into account a variety of practical aspects including the energy consumption, user distribution, and weather conditions. During peak hours without wind, the foldable array can spread to form a larger composite array of various shapes in alignment with the distribution of user clusters. 

It should be noted that the above architectures may share certain similarities and can be classified under the broad category of joint 3D orientation and 3D movement design. However, they are designed for application in different scenarios based on their specific features, thereby facilitating practical implementation, as previously discussed.

It is also worth noting that all the above array-level MAs can also be implemented in a hybrid manner with the conventional FPAs at the current cellular BSs, as shown in Fig. \ref{MA_FIG}, thereby enhancing their versatility. For instance, an auxiliary sliding array can be added to the existing tri-sector BS architecture to offload traffic or eliminate coverage gaps in each sector. Similar hybrid architecture can also be applied to the rotatable, turnable, and foldable arrays. As illustrated in Fig. \ref{MA_FIG}, there is a fixed circular, cross-shaped, and rectangular sub-array at the center of the rotatable, turnable, and foldable array, respectively. These fixed sub-arrays are essential for transmitting key control signals and receiving user access requests to discover new users. Based on user distributions, their movable sub-arrays can be adjusted accordingly to ensure optimal communication performance.

\subsection{Dual-Scale MAs}
In the above architectures, the movement range of MAs is restricted to a relatively small region at the transmitter or receiver. This may prevent MAs from fully enhancing wireless channel conditions. For instance, in complex environments such as urban areas, signal transmission can be frequently disrupted by scattered and dense obstacles. Such LoS blockage issues cannot be resolved by antenna movement within a small region due to its much smaller size compared to the dominant environmental scatterers.

To address this challenge, we propose a more general concept of dual-scale MAs by mounting the MAs on mobile platforms, such as unmanned aerial vehicles (UAVs) and terrestrial vehicles, thereby reaping the combined benefits of both large- and small-scale channel reconfiguration. For example, the locations of the mobile platforms can be optimized first to ensure strong path gains with user clusters, followed by optimizing the locations and orientations of the MAs to maximize their coverage performance. Additionally, UAV-mounted MAs can help extend communication and sensing coverage beyond the existing cellular network by establishing LoS paths with numerous terrestrial users and adjusting the MA locations and orientations accordingly. However, the adoption of dual-scale MAs introduces several new challenges. For example, the combination of large-scale mobility and local antenna adjustments requires advanced algorithms for location and orientation optimization. Furthermore, the performance of UAV-mounted MAs may be affected by wind disturbances, resulting in inaccurate positions/rotational angles and necessitating more robust designs.
\begin{figure*}
    \centering
    \includegraphics[width=0.98\linewidth]{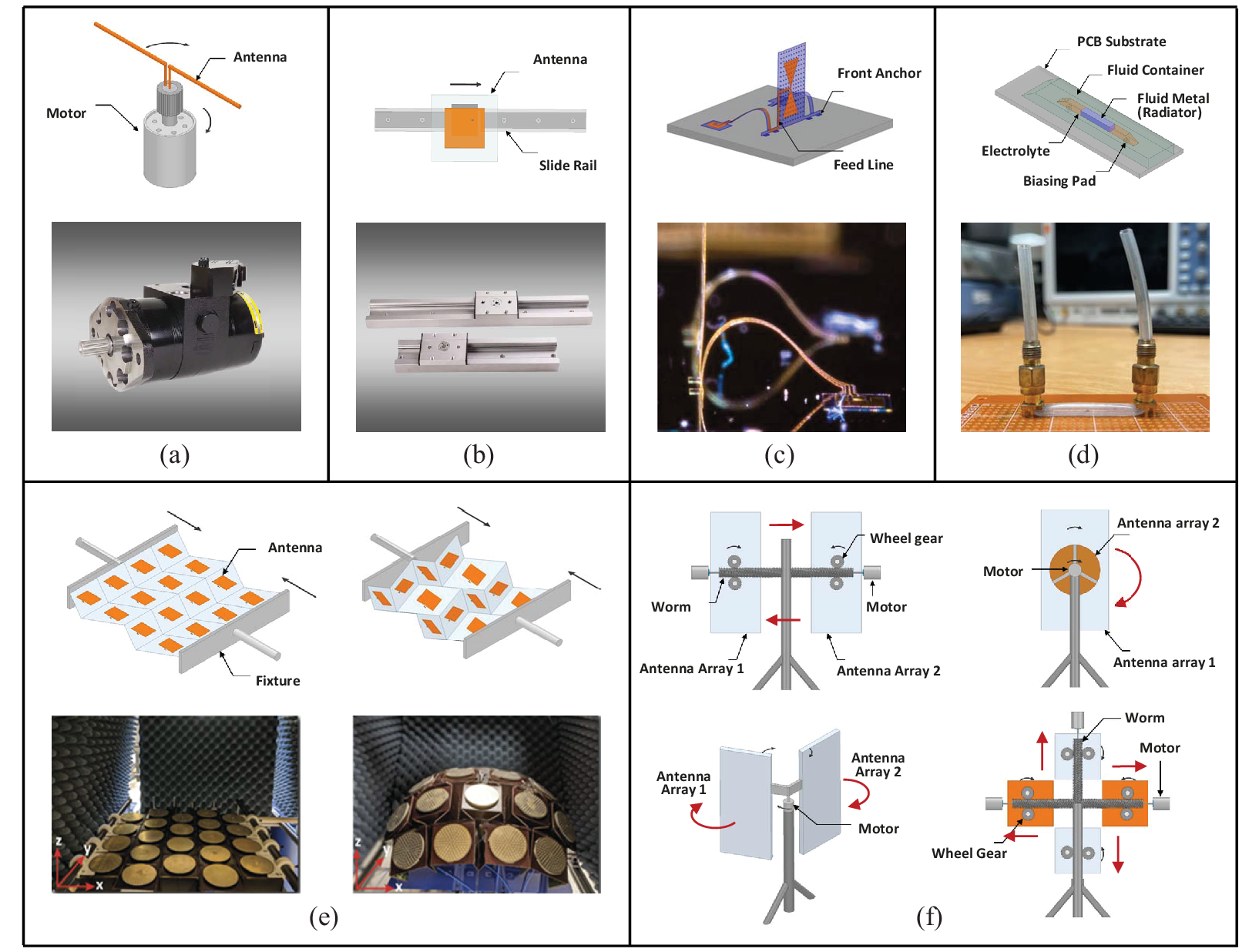}
    \caption{Mechanically driven MAs by (a) a motor, (b) a slide ray, (c) MEMS, (d) liquid fluidity, and (e) internal structure transformation, and (f) mechanically driven antenna arrays.}
    \label{mechanical}
\end{figure*}

\section{Practical Implementation of MAs}\label{imp}
In this section, we present some viable methods to implement the aforementioned MA architectures in practice. These approaches can be broadly categorized into two types: mechanically driven and electronically driven.

\subsection{Mechanically Driven MAs}
\subsubsection{External Machinery} One straightforward way to move or rotate antennas as desired is by using external mechanical structures equipped with actuators that convert control signals and energy into torque or displacement. Some possible actuators for MAs include electric motors and micro-electromechanical systems (MEMS). As illustrated in Fig. \ref{mechanical}(a) and \ref{mechanical}(b), electric motors can achieve 1D rotation and 2D movement of antennas by either mounting the antennas directly on the motor shaft \cite{SA1} or by driving a lead screw with the motor shaft \cite{SA2}. MEMS, on the other hand, can facilitate antenna rotation through voltage or current actuation \cite{SA3}, as shown in Fig. \ref{mechanical}(c). However, to fully exploit the DoFs in the transmit/receive region for MAs, the mechanical flexibility of these architectures needs further improvement. Additionally, efficiently and simultaneously moving or rotating multiple antennas in an energy-efficient and low-latency manner is a critical issue that must be addressed.

\subsubsection{Liquid Fluidity} Liquid fluidity has also been considered a promising approach for implementing MAs (in which case MAs become equivalent to fluid antennas) by employing the continuous electrowetting (CEW) techniques. In this approach, a fluid metal is employed as an antenna within a liquid container, as shown in Fig. \ref{mechanical}(d). By applying a voltage to two electrodes, charge redistribution occurs on the surface of the fluid metal, altering the surface tension and causing the liquid metal to move within the container via Marangoni forces. A prototype of fluid antennas, as proposed in \cite{BU1}, is shown in Fig. \ref{mechanical}(d). The practical implementation of fluid antennas faces challenges in selecting a suitable liquid material, which must meet various criteria such as cost, safety, physical and chemical stability, melting point, and viscosity. Additionally, properties that directly affect antenna performance, such as permittivity, permeability, conductivity, and loss, must also be considered. Moreover, how to achieve functionality of antenna rotation with fluid antennas is also an interesting problem worthy of further investigation.

\subsubsection{Internal Structure Transformation} Deployable structures, a type of transformable structure that can expand and contract, have led to the development of deployable antenna arrays \cite{BU2}. These arrays leverage internal mechanical structures to alter their geometry (e.g., shape and position) for various purposes, such as space storage optimization and performance reconfiguration. The geometry of deployable antenna arrays is often based on origami folding techniques, as illustrated in Fig. \ref{mechanical}(e). By changing the antenna geometry, the diversity of electromagnetic fields in the spatial domain can be enhanced, enabling capabilities like adaptive imaging and beam steering \cite{SA9}. Compared to external machinery, deployable antenna arrays offer a simpler and more cost-effective way to implement MAs. However, the range of movement/rotation of deployable antenna arrays is limited by the inherent mechanical properties of the deployable structure. Consequently, the flexibility in antenna movement and rotation is reduced compared to MAs using external machinery.

\subsubsection{Application to Array-Level MAs} Note that the deployable structure can be directly applied to change the position/rotation of antenna arrays. The external machinery can also be used to drive the movement of multiple antenna arrays. As shown in Fig. \ref{mechanical}(f), a possible way is to integrate the worm and wheel gear sets into the array and the fixture. The motor drives the rotation of the worm, thereby producing the movement/rotation of the array. Note that large conveyors may be required when moving a large antenna array, such as vehicles and conveyor belts. The liquid fluidity faces more difficulty to mobilize multiple antenna arrays due to a large-scale use of liquid material, which needs to meet a wide range of criteria as previously discussed.

\subsection{Electronically Driven Equivalent MAs}
Unlike the previous approaches achieving MA through physical movements and rotation, equivalent antenna movement and rotation can also be implemented in an electronic manner.

\subsubsection{Dual-Mode Antennas} First, we introduce the concept of an antenna's phase center. The phase center is defined as the effective origin of radiation, from which spherical waves emanate with constant phase fronts in the far-field region. Moving the antenna results in a shift of its phase center. Therefore, the equivalent physical movement of an antenna can be achieved by adjusting its phase-center location. This is feasible for e.g., dual-mode patch antennas \cite{SB1}. Specifically, the phase center of a circular patch antenna aligns with its physical center when operating in a single mode. However, by exciting multiple modes concurrently, the phase center can be displaced from the patch's center. For example, a stacked dual-mode circular patch antenna configuration that excites the TM11 and TM21 modes can achieve this equivalent movement, as shown in Fig. \ref{ED}(a). This approach allows a uniformly-spaced antenna array to be transformed into a non-uniform configuration without any physical adjustments.

\subsubsection{Dense Array Antennas} 
\begin{figure*}
    \centering
    \includegraphics[width=0.98\linewidth]{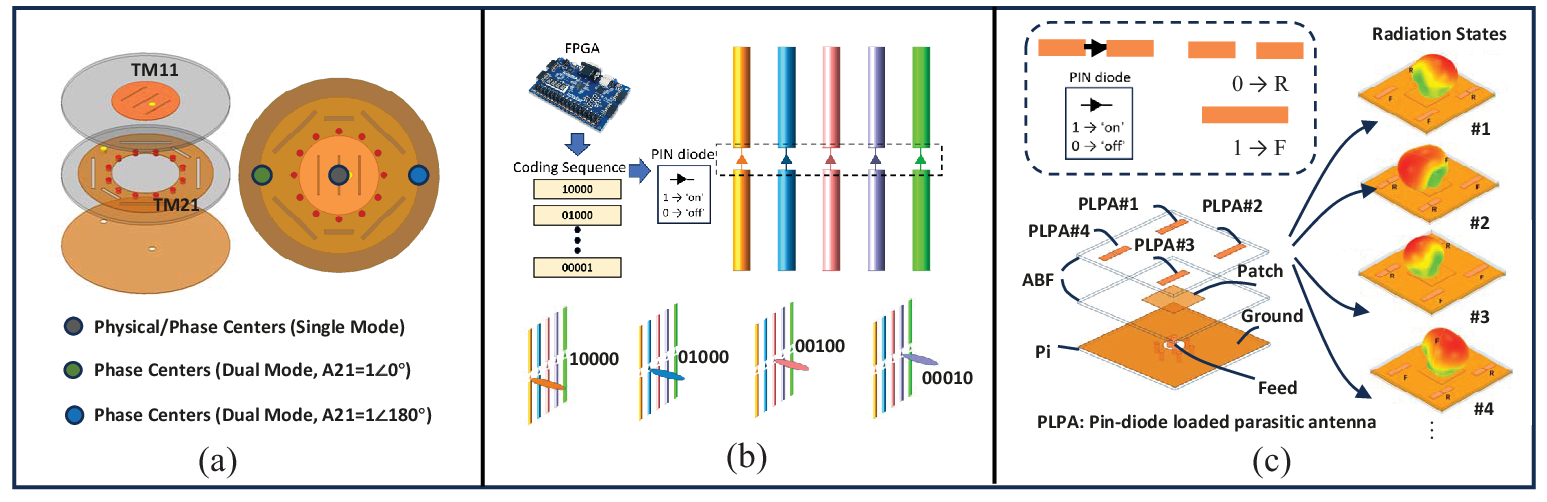}
    \caption{Electronically driven MAs: (a) dual-mode antennas, (b) dense antenna array, and (c) orientation-reconfigurable antennas.}
    \label{ED}
\end{figure*}

Equivalent movement of antenna positions can also be achieved by the use of dense array antennas. In particular, a multitude of antennas are densely deployed within the transmit/receive region each being loaded with a reconfigurable device, e.g., a PIN diode, shown in Fig. \ref{ED}(b). By controlling the states of these reconfigurable devices (e.g., via a coding sequence from an FPGA), different sets of antennas can be activated. For example, if the initial coding sequence is ``10000'', only the first antenna would be activated while other antennas remain inactive. Similarly, the active antenna can be switched from the first antenna to the last antenna by changing the coding sequence, thereby approximating the antenna movement in the continuous space. Notably, as shown in our previous work\cite{EMA1}, a moderate antenna spacing (e.g., one-sixth wavelength) suffices to achieve comparable performance to continuous antenna movement. As such, dense array antennas turn out to be an efficient approach for the practical implementation of MAs, especially in the case that a small movement space and dimension are desired.

\subsubsection{Orientation-Reconfigurable Antennas} Orientation-reconfigurable antennas can achieve equivalent antenna rotation by adjusting the pointing direction of an antenna beam without physically rotating the antenna. As shown in Fig. \ref{ED}(c), a patch antenna is loaded by four pin-diode loaded parasitic antennas (PLPAs). The working state of each loaded pin diode can be controlled by an FPGA to control the direct current (DC) bias and switched between ``on'' and ``off'' states. The PLPAs work as a director when the pin diode is turned off; otherwise, it acts as a reflector \cite{SB2}. As a result, diverse radiation beams pointing to different directions can be generated by controlling the working states of the four loaded PLPAs, ultimately achieving equivalent antenna rotation with quick response and high accuracy.

\subsection{Comparisons}
Mechanically and electrically driven MAs each possess distinct advantages and limitations in terms of various practical aspects.

\subsubsection{Speed of Reaction and Movement} Due to the limited flexibility of mechanical control, mechanically driven MAs typically exhibit slower reaction speeds. Particularly, in the case of global antenna element/array movement, the movement of some elements or arrays may be constrained by the positions of others to avoid collisions, further reducing their movement speed. In contrast, electrically driven MAs eliminate the need for physical movement of antenna elements or arrays, resulting in faster response time to achieve the desired antenna geometry and smaller waiting time where the array cannot be used.

\subsubsection{Range of Movement} The movement range of mechanically driven MAs depends mainly on the adjustable range of the control system. The mature design process and straightforward principles of mechanical systems make it easier to achieve full-coverage applications. In contrast, electrically driven MAs typically achieve equivalent movement or rotation through multiple reconfigurable radiation states. Although their response is faster, the limited number of tunable states often makes it more challenging to achieve a broad movement range. 
\begin{figure*}
    \centering
    \includegraphics[width=0.98\textwidth]{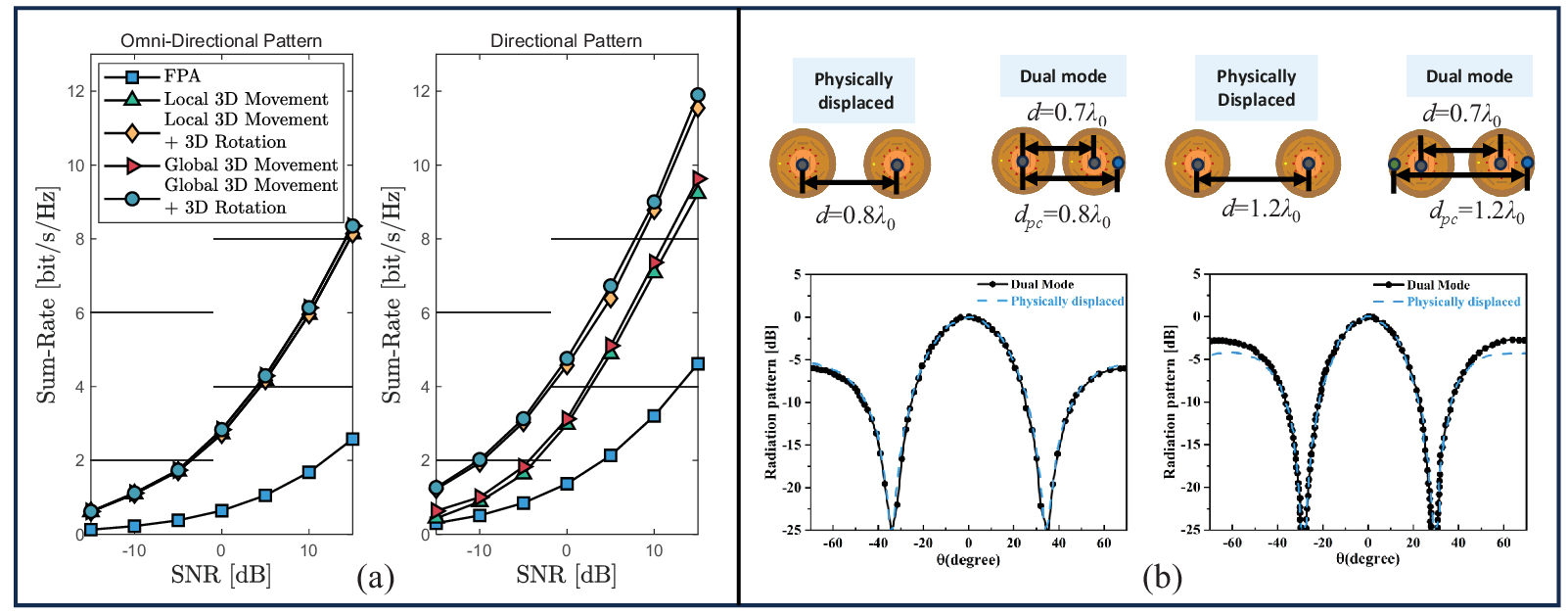}
    \caption{(a) Sum-rate performance versus transmit SNR by omnidirectional and directional MAs; (b) radiation patterns of the dual-mode two-element antenna array under different setups.}\label{EMO} 
    \vspace{-12pt}
\end{figure*}

\subsubsection{Movement Resolution} Mechanically driven MAs can achieve a high movement resolution using precision mechanical drive systems and high-resolution encoders. However, environmental factors such as temperature, humidity, and vibration can degrade this accuracy over time. Additionally, wear and tear from prolonged use can further reduce their movement accuracy and reliability. In contrast, the movement resolution of electrically driven MAs depends on the precision of electronic technology and control algorithms. Variations in electronic component performance and temperature changes can affect their accuracy. However, since electrically driven MAs do not require physical movement of antennas, they avoid accuracy degradation from mechanical wear and tear. This advantage enhances the system's long-term reliability and stability.

\subsubsection{Cost and Maintenance} The implementation cost of mechanically driven MAs varies based on application requirements and performance specifications. Generally, their cost is relatively low due to a mature design and manufacturing process, along with simple installation and commissioning. However, mechanically driven MAs are prone to wear and tear over time, necessitating regular maintenance and calibration to maintain accuracy and reliability, which can increase operational costs and downtime. In contrast, electrically driven MAs have higher implementation costs due to complex design, stringent quality control, and more involved installation and commissioning processes. However, the absence of wear-prone mechanical parts results in less frequent maintenance, reducing downtime and operational interruptions.

It follows from the above that the choice between these two types of MAs should depend on specific application requirements, budgetary constraints, and technical considerations. For instance, electrically driven MAs are preferable in delay-sensitive environments such as vehicle and satellite communications, whereas mechanically driven MAs are more suitable for delay-tolerant applications like the Internet of Things (IoT) and smart home systems.

\section{Performance Evaluation}
Numerical results are presented in this section to demonstrate the effectiveness of the proposed MA architectures and implementation methods. 

First, we show the performance of element-level MAs based on the following setup, i.e., a BS equipped with 4 MAs serves 4 users, each equipped with a single FPA. In particular, each MA element at the BS can be either fixed, locally moved or globally moved, with or without rotation capability. In addition, each MA can be directional (with the radiation pattern set based on the 3GPP technical report 38.901\footnote{The technical report is available online at \url{www.3gpp.org/DynaReport/38901.htm}.}) or omni-directional. We aim to jointly optimize the 3D position and 3D orientation of each MA under the zero-forcing (ZF) precoding. We solve this rate maximization problem using Bayesian optimization\footnote{Readers can refer to an online tutorial at \url{https://arxiv.org/pdf/1807.02811} for more details about this algorithm.} and show the users' sum rate versus the transmit signal-to-noise ratio (SNR) in Fig. \ref{EMO}(a) for omni-directional and directional MAs, respectively. First, it is observed from Fig. \ref{EMO}(a) that the schemes with MAs using omni-directional patterns yield significant performance gain ($220\%$ higher) over that with FPAs. Nonetheless, joint 3D movement and 3D rotation is observed to achieve the same sum-rate as 3D movement only for omni-direction MAs, with either local or global movement. This is expected, as the radiation pattern of omni-directional antennas is regardless of their rotational angles. Moreover, it is observed from Fig. \ref{EMO}(a) that local antenna movement yields comparable performance to global antenna movement, thus enabling more cost-effective implementation of the MA technology. Most of the above observations for omni-directional MAs are similarly made for directional MAs in Fig. \ref{EMO}(a). However, unlike omni-directional MAs, joint 3D movement and 3D rotation of directional MAs is observed to yield much better performance than their 3D movement only. The reason is that the radiation pattern of directional antennas critically depends on their rotational angles.

Next, we evaluate the feasibility of using a two-element dual-mode (TM11 and TM21 modes) antenna array for equivalent antenna movement in Fig. \ref{EMO}(b), under two different setups. In these setups, different excitation schemes are applied to alter the distance between the physical centers of the two dual-mode antennas, denoted as $d_{pc}$. We set $d_{pc}=0.8\lambda$ and $d_{pc}=1.2\lambda$ in the first and second setups, respectively, with $\lambda$ denoting the wavelength. As observed from Fig. \ref{EMO}(b), the radiation pattern of the dual-mode antenna array closely resembles that of a conventional antenna array with a physical element spacing of $d = d_{pc}$ in both setups. This suggests that dual-mode antennas can be a cost-effective approach to achieve equivalent antenna movement without the need for physical adjustments.

\section{Conclusion} 
In this article, we provided an overview of general architectures and implementation methods for MA-aided communication systems. The proposed architectures fully exploit the extra DoFs in e.g., 3D rotation and large-scale movement while being tailored to specific communication requirements and scenarios, thereby effectively balancing overall cost and communication performance. Additionally, we presented two types of implementation methods for the proposed MA architectures that utilize mechanical and electronic controls, respectively. Numerical results demonstrated that electronic control can achieve equivalent antenna movements without any physical displacement of the antennas. 

It is worth noting that, in addition to the implementation methods discussed in this article, there may be other potential mechanical and electrical approaches due to the long-standing advancements in reconfigurable antenna technologies. For example, the mechanical properties of certain flexible materials, such as shape memory polymers, can be altered by external stimuli like light, which could be utilized to modify antenna geometry in specific scenarios. It is hoped that this article can serve as a stepping stone for the integration and widespread adoption of MAs in future wireless networks.

\section*{Acknowledgement}
The authors would like to thank Dr. Liwei Zhao and Dr. Yuhan Fan from the National University of Singapore for their assistance in Section \ref{imp}.

\bibliographystyle{IEEEtran}
\bibliography{Reference}
\end{document}